# Strong coupling regimes of an organic exciton mirror in a microcavity

*Christoph Bennenhei[1], Lukas Lackner[1], Moritz Gittinger[1], Falk Eilenberger[2], Marvin F. Schumacher,[3] Arne Lützen[3], Ivan Shelykh[4], Christoph Lienau[1], Martin Esmann[1], Christian Schneider[1]*

[1]*Institute of Physics, Fakultät V, Carl von Ossietzky Universität Oldenburg, 26129 Oldenburg, Germany*

[2]*Institute of Applied Physics, Abbe Center of Photonics, Friedrich Schiller University Jena, 07743 Jena, Germany; Fraunhofer-Institute for Applied Optics and Precision Engineering IOF, 07743 Jena, Germany; Max-Planck-School of Photonics, 07743 Jena, Germany*

[3]*Kekulé-Institute for Organic Chemistry and Biochemistry, University of Bonn, 53121 Bonn, Germany*

[4]*University of Iceland, Science Institute, Reykjavik, Iceland*

**ABSTRACT**

The coherent, periodic energy transfer between light- and matter excitations characterizes the strong coupling regime of cavity exciton-polaritons, resulting, in the simplest case, in a Rabi-doublet in the spectral domain. We demonstrate a peculiar regime of strong light-matter coupling, which arises when photonic cavity modes couple to an ultra-thin excitonic mirror. We embed a 12 nm J-aggregated thin film in an open microcavity and tune the coupling strength from weak to the onset of ultrastrong coupling. At resonance, the excitonic mirror selectively changes dielectric to metallic field boundary conditions adding a 2π phase, which links optical cavity modes of different order. Our work gives an exciting perspective to ultra-fast cavity switches and photonic devices based on excitonic optical elements.

## INTRODUCTION

Strong coupling of excitons to cavity-confined photons leads to the formation of cavity polaritons, which are bosonic light-matter hybrid states. These excitations have become a matter of extensive research interest due to their ability to display intriguing thermodynamic behavior, including phase transitions to a bosonic condensate at elevated temperatures [1,2], Berezinskii-Kosterlitz-Thouless physics [3,4], superfluidity and vorticity [5,6], as well as non-linear quantum optical phenomena [7,8] and low-threshold lasing [9]. With the rise of a larger plethora of quantum materials displaying intriguing excitonic phenomena, the portfolio of exciton-polaritons was vastly extended towards room-temperature exciton-cavity systems, magneto polaritonics and van-der-Waals exciton-polaritons [10,11].

Organic molecules hosting tightly bound Frenkel excitons are a particularly promising and easily accessible platform for the realization of room-temperature polaritonic systems with extraordinarily large Rabi-splittings [12–16]. The exciton Bohr radius of roughly the size of a single molecule yields binding energies on the order of 1 eV, making these excitons stable at ambient conditions. At the same time, the optical properties of certain J-aggregated organic layers [17–21] feature a strongly reduced inhomogeneous broadening [22], small dephasing [23] and a highly dispersive dielectric function, which features a pronounced exciton-reflectivity even for layers with thicknesses smaller than 10 nm [24].

We employ such a system, based on a J-aggregated squaraine layer in a spectrally tunable open cavity [25–31], to explore the fingerprint features of the resulting strong coupling transitions at room-temperature. The very large oscillator strength of the layer allows us to resolve a Rabi-splitting approaching 10% of the exciton energy, which consistently reduces as we increase the cavity length. However, contrasting a conventional strong-to-weak coupling transition, we observe that at large cavity lengths, the mirror-like behavior of the excitons [32–35] in our organic layer starts to dictate the properties of the systems. In this regime, the excitons selectively enforce metallic field boundary conditions in the cavity over a specific range of energies even in the limit where the excitonic material makes up below 0.1% of the cavity volume.

The structure of our samples is sketched in Fig. 1(a). It is based on an open optical cavity, which consists of a bottom distributed Bragg reflector (DBR), a thin film of squaraine and a gold top mirror. The DBR consists of 10 $SiO_2/TiO_2$ pairs, designed for a central wavelength of the stopband at 750 nm and a field maximum in the active material. We mount the DBR and a (100 μm)$^3$ glass mesa coated with a 45 nm thick gold mirror on stacks of nano-positioners as described in Ref. [26], such that the mirrors form a cavity separated by an air gap. The resulting open cavity yields full spectral tunability via the length of the air gap. We investigate two samples of the J-aggregated squaraine (*S*,*S*)-ProSQ-C16 (see Fig. 1(b)), synthesized as described in Ref. [36], which is spin-coated and annealed on the bottom DBR, following the process reported in Ref. [22]: The primary sample of this investigation has a 12 nm thick squaraine layer and the secondary sample features a much thinner layer, which we determine to range on the order of ~1 nm via transfer-matrix calculations.

Fig. 1(c) compiles optical properties of the excitons in the 12 nm thin squaraine layer, recorded via photoluminescence (PL) and transmission spectroscopy, acquired on a glass substrate. Strong exciton PL from the J-band is observed at 1.59 eV along with a characteristic reduction in optical transmission. The primarily J-type of the aggregation is highlighted by the significantly less pronounced response of the H-band, observed in transmission. We point out that the high oscillator strength leads to a dielectric response of the thin film, which is characterized by a negative real part of the dielectric function for 1.6-1.7 eV [24] (see Fig. 4(f)). This is a feature that is commonly seen in the dielectric response of lattice vibrations as a phonon Reststrahlen band in the infrared [37]. It is far less common for excitons in the visible part of the spectrum. This quasi-metallic behavior impedes optical propagation through the material as it turns the propagating waves in the layer into evanescent waves.

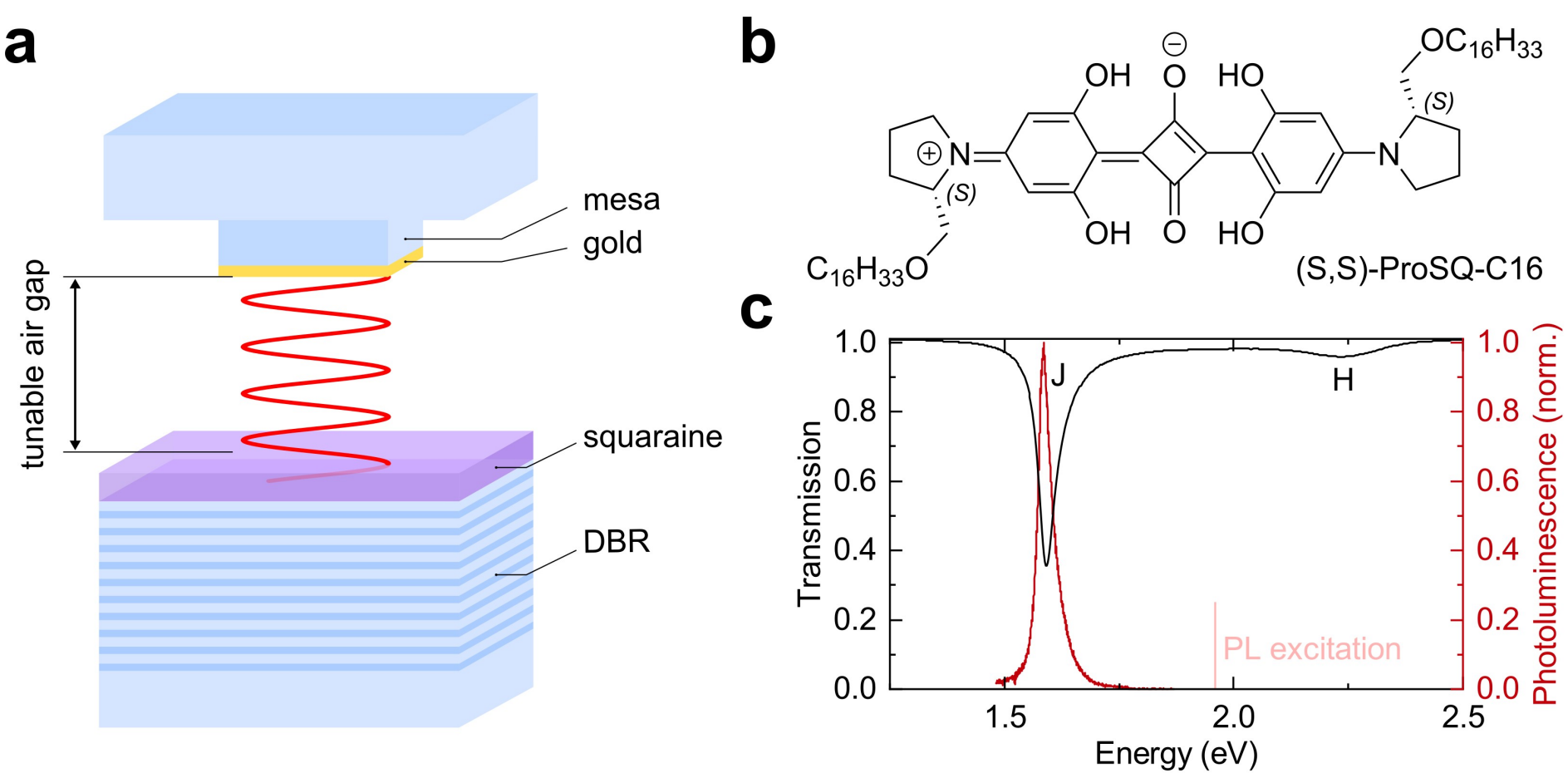

**Figure 1.** (a) Sketch of the tunable open microcavity with a J-aggregated squaraine thin film as the active material positioned in the field maximum of the cavity. (b) Molecular structure of (S,S)-ProSQ-C16. (c) Room-temperature photoluminescence and transmission spectra of a reference 12 nm thin film on a glass substrate.

We investigate the strong coupling regime between excitons in the squaraine film and the photonic resonance of our cavity via momentum-resolved white light reflection spectroscopy (a schematic of the experimental setup is shown in Supplementary Figure S1). In Fig. 2, we depict a series of dispersion relations from our device, which we recorded while tuning the air gap distance from 620 nm to 430 nm. In the 620 nm case (a), the spectrum is characterized by a set of parabolic dispersive branches (UP, LP) with the one around 1.5 eV being the negatively detuned lower polariton resonance (LP). As the cavity length is reduced (b,c), we see that this branch shifts towards the exciton transition at 1.59 eV and its effective mass strongly increases, until the branch becomes almost dispersionless at an air gap distance of 430 nm (d).

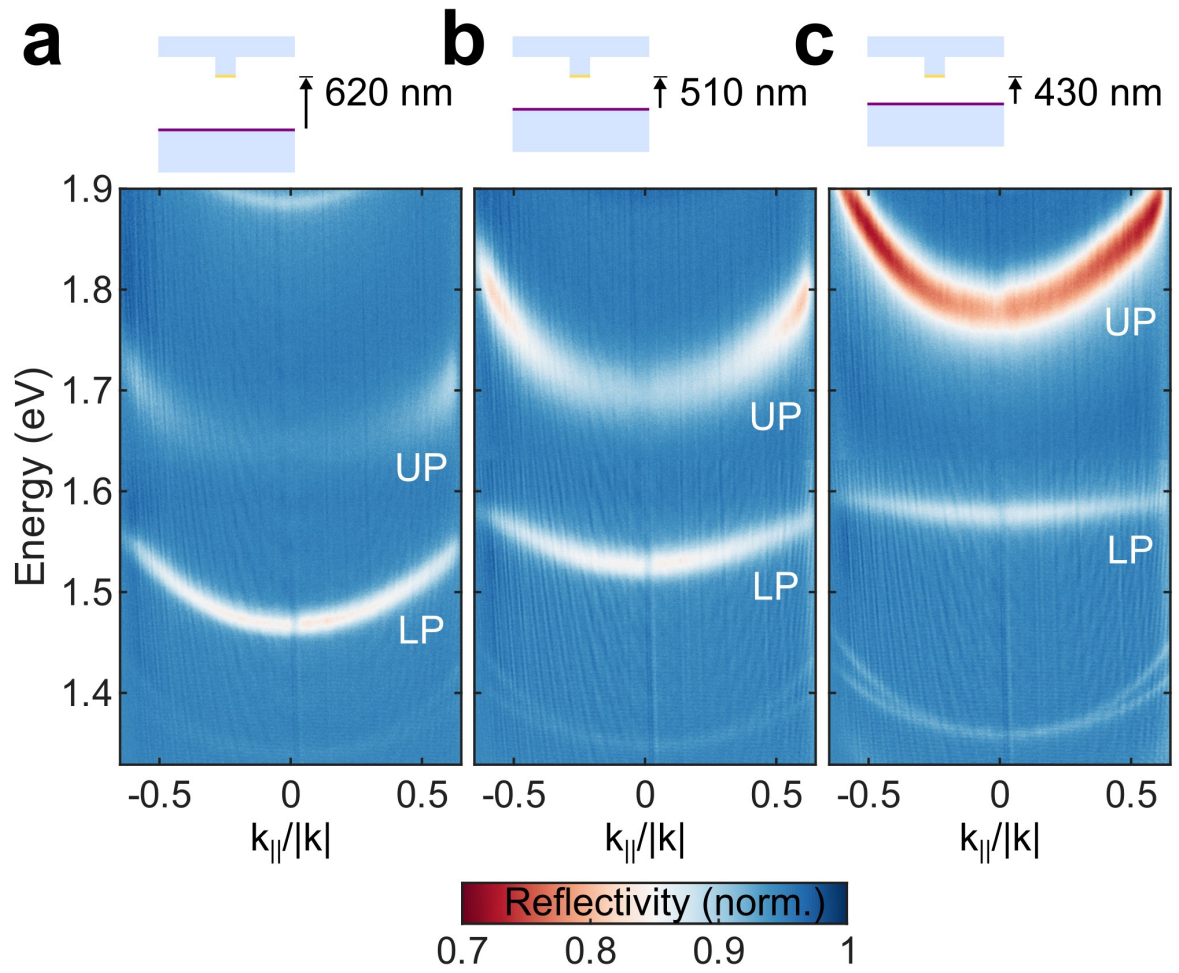

**Figure 2.** Momentum resolved white light reflectivity spectra for different exemplary air gap distances (as indicated by the sketches in the top row) from a microcavity with a 12 nm squaraine film. Clear polariton (UP,LP) behavior is observed by the formation of the characteristic anti-crossing and the flattening of the lower polariton branch (LP) in the vicinity of the exciton at 1.59 eV.

The anti-crossing behavior can be even better captured in $k_{\|} = 0$ cuts of the reflectivity spectra, which we plot as a function of the air gap distance in Fig. 3(a). Here, we tune consecutive longitudinal cavity modes through the exciton resonance, which all feature the same characteristic bending behavior on resonance, which is accompanied by a reduction of the reflection contrast. In a phenomenological coupled oscillator picture with one collective excitonic mode coupled to one photonic cavity mode, the emergence of exciton-polaritons is characterized by a full anti-crossing [38,39]. In our sample however, due to the very large Rabi-splitting, comparable in magnitude to the longitudinal mode spacing, consecutive photonic modes collectively couple to the exciton, which transforms the anti-crossing into the characteristic bending as the cavity photon couples resonantly to the exciton. This behavior is quantitatively reproduced by the spectrally resolved transfer-matrix calculation, shown in Fig. 3(b) and in Supplementary Sections S2 and S3. The size of the Rabi-splitting is analyzed for every optical mode from a much longer cavity scan (the full scan is shown in Supplementary Figure S2) as they couple to the excitonic oscillator, and is plotted as a function of the absolute length of the cavity air gap in Fig. 3(c). The typical behavior of a polaritonic system with tunable cavity length is captured in our system: Increasing the cavity length reduces the Rabi-splitting, as the dipole coupling strength scales with the inverse square root of the effective length of the cavity. We approximate the behavior as $\Omega = \frac{A}{\sqrt{L_{cav}}}$, (red curve in Fig. 3 (c)) where $\Omega$ is the Rabi-splitting, $L_{cav}$ is the effective length of the cavity, and A is a free parameter proportional to the strength of the exciton oscillator. At a cavity length ~11 μm the system transitions from the weak to the strong coupling regime, see Supplementary Section S4. We stress that for the shortest cavity length the magnitude of the Rabi-splitting exceeds 10% of the exciton energy, thus our device operates at the onset of the ultrastrong coupling regime [40,41].

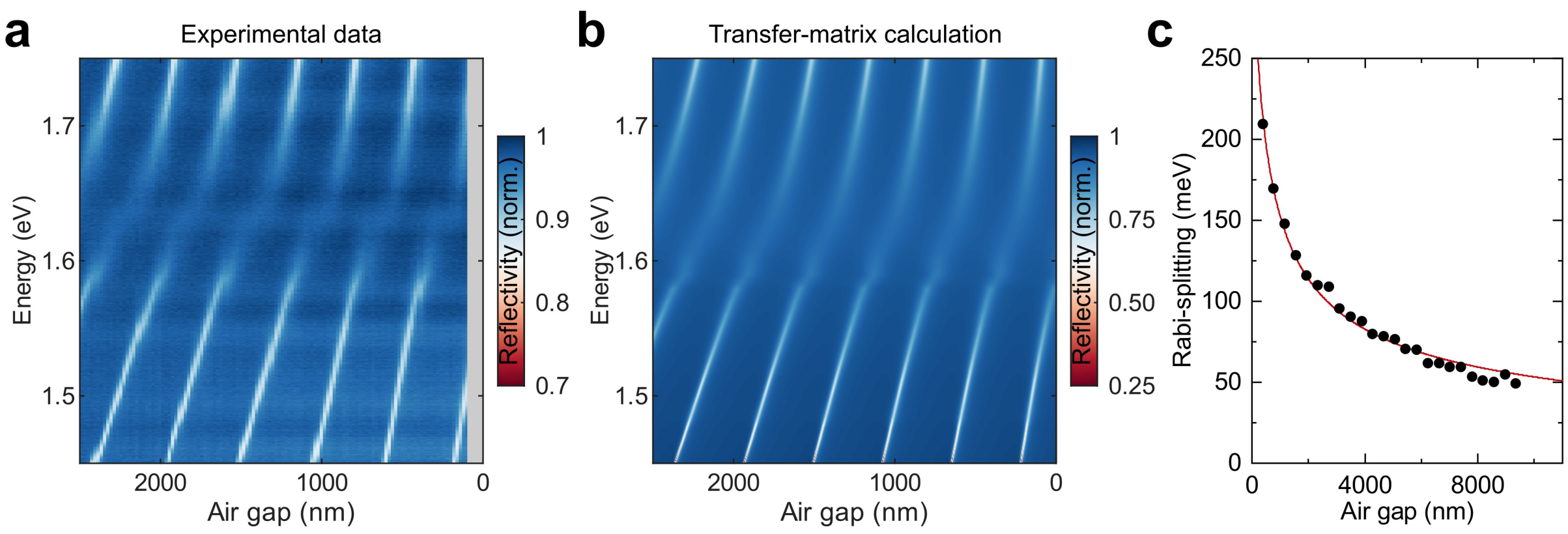

**Figure 3.** (a) Normal incidence reflectivity spectra ($k_{\parallel} = 0$ sections) recorded as a function of the air gap of the cavity with a 12 nm squaraine film. The gray area marks contact between the mirrors, inhibiting a closer approach. Anti-crossings occur at the exciton energy of 1.59 eV (b) Transfer-matrix calculations corresponding to the experimental data shown in panel (a). (c) Extracted Rabi-splitting $\Omega$ as a function of the air gap (symbols) and fitted curve following $\Omega = \frac{A}{\sqrt{L_{cav}}}$.

In the following, we will now turn our attention to a peculiar coupling scenario which results from the interplay of the very large Rabi-splitting in our device, in conjunction with the impeded propagation of light through the thin squaraine layer, acting as an embedded excitonic mirror in our cavity:

In Fig. 4(a), we first depict a schematic of a standard exciton-polariton system featuring a modest Rabi-splitting, as the cavity length is tuned over a wide range. This scenario can be observed for a thinner (~1 nm) layer of squaraine in our cavity. It is confirmed by experimental data which are shown in Fig. 4(b). Therein, we plot $k_{\parallel} = 0$ cuts of recorded reflectivity spectra as a function of the cavity length and deduce a Rabi-splitting of 60 meV at the shortest cavity length (air gap distance of 220 nm). The full set of experimental data and corresponding numerical calculations is shown in Supplementary Figure S3.

For the sake of clarity, we label in the schematic in Fig. 4 (a) each longitudinal mode with an index $n$. We see that in this standard scenario, the Rabi-splitting reduces for each longitudinal mode, until the weak coupling transition is reached, where longitudinal modes with ascending mode numbers each cross the exciton resonance as a straight line when plotted as a function of wavelength. In energy representation, a straight line remains a good approximation to the cavity dispersion over small energy intervals.

In Fig. 4(c), we show a schematic for a non-standard strong coupling case, in the excitonic mirror regime, which contrasts the scenario depicted in Fig. 4(a,b). For shortest cavity lengths, the large coupling strength results in a characteristic S-like 'bending' of the optical resonance, which can prominently be seen in the experiment. As the cavity length increases, the curvature of the modes near the exciton resonance slowly disappears. This is in contrast to the standard picture of a strong-to-weak coupling transition, where the upper and lower polariton branches attain an increasing curvature in the vicinity of the exciton resonance, as the Rabi gap closes. This is shown experimentally in Fig 4(d) and in Supplementary Figure S2 in comparison to a corresponding transfer-matrix simulation. Experimentally, we observe the straightening of the optical resonances as they shift through the exciton oscillator. Note that a difference in slope is expected on the two sides of the exciton since we effectively connect modes with slopes proportional to n and n+1 in the empty cavity. This difference, however, diminishes for long cavities.

As a consequence, for large cavity lengths, the optical mode, which crosses the exciton energy, in fact changes its mode index during the crossing, which contrasts the standard weak coupling case. The observed linking of modes of different mode order as a consequence of the excitonic mirror persists in both the strong and weak coupling limit (see Supplementary Figures S4 and S5). As we demonstrate via transfer-matrix calculations in Supplementary Sections S2 and S5, the mode index directly corresponds to the number of field anti-nodes for the assigned optical resonance. In the excitonic mirror regime this changes by integer 1 for an apparent straight mode when crossing the exciton resonance. As depicted in Fig 4(f), this observation can effectively be viewed as a direct consequence of a phase change by $2\pi$ in reflection phase of the excitonic mirror compared to a bare DBR when crossing the exciton resonance. Such a $2\pi$-phase change differs from the π-change introduced by the exciton layer alone in the commonly employed Lorentz oscillator model for the exciton. Similar 2π-phase changes have recently been observed, for example, in perovskite nanoparticles [42], but have not been reported in polaritonic cavities.

We now establish a scenario in which the thin organic material separates the bottom DBR from the air gap, which is terminated by the upper gold mirror (Fig. 4(e)). As the thin squaraine layer acts as a spectrally selective mirror, it induces a phase shift of the optical mode on the exciton resonance. This is verified by the numerical calculation in Fig. 4(f) showing the optical reflection phases of a single, bare DBR (black), DBR with 1 nm, corresponding to the 'standard' case (blue) and 12 nm thick layer of squaraine, corresponding to the excitonic mirror regime (red). See Supplementary Figure S7 for a thickness sweep. For energies well above the exciton resonance, the added squaraine essentially does not change the lower DBR's reflection phase and cavity round trip phase. On resonance, the 12 nm thick squaraine acts as an ultra-thin excitonic mirror and enforces a field node of the optical mode profile at its surface, i.e. the reflection phase changes by $\pi$ (Fig. 4(e, middle)). Below the exciton resonance, the dielectric mirror again enforces an anti-node resulting in an overall change in round-trip phase of $2\pi$ across the exciton resonance and thus in the unconventional change in photonic mode number by a full integer in the long cavity limit, which is an unambiguous sign of the excitonic mirror physics occurring in our device. Interestingly, for the 1 nm thin squaraine, the optical penetration depth is large enough for the DBR to dominate the overall reflection phase of the lower cavity section and thus, the phase jump is absent, restoring the conventional weak-coupling limit.

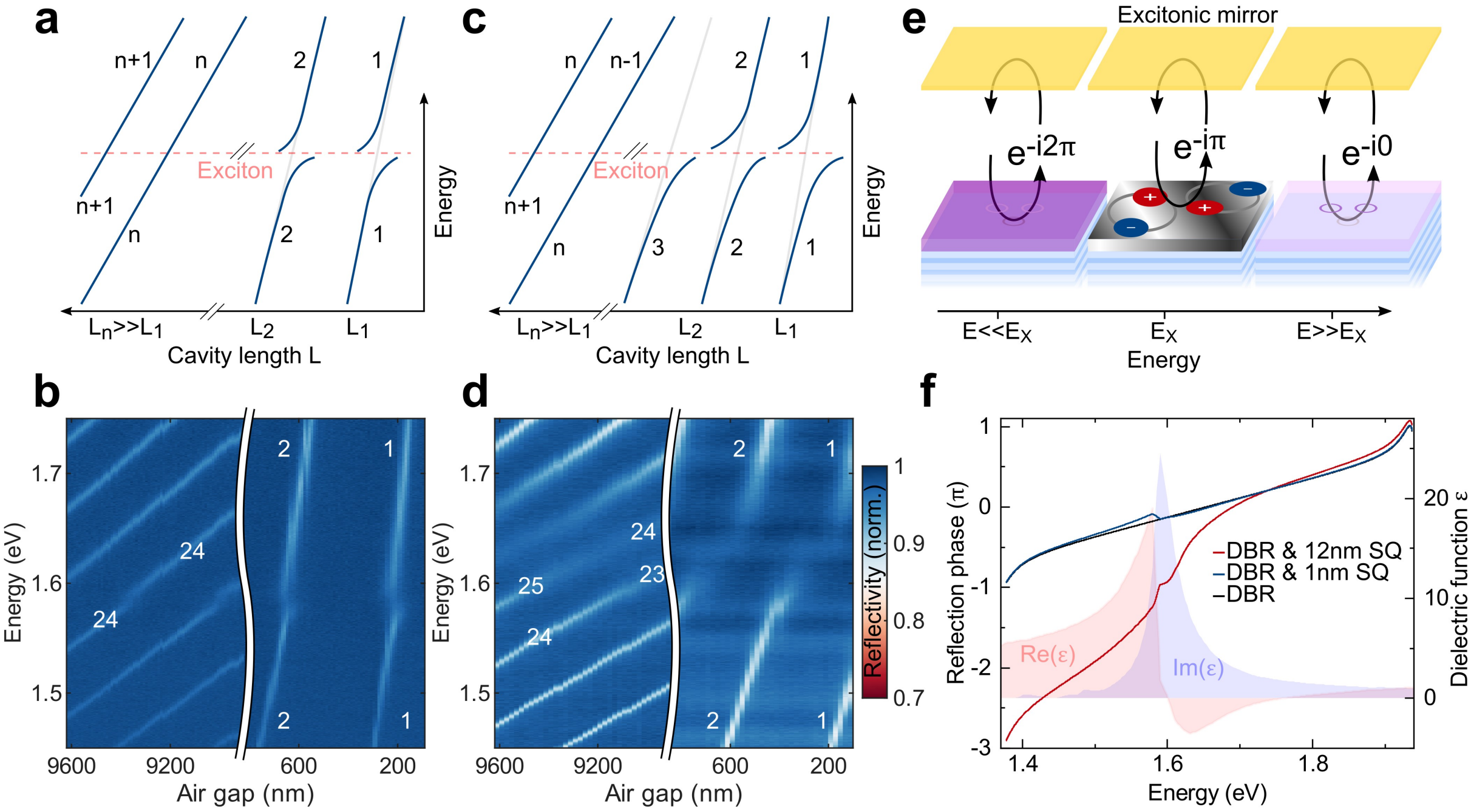

**Figure 4.** (a) Schematic and (b) experimental data of the transition from strong to weak coupling for a device with a ~1 nm squaraine layer. (c) Schematic and (d) experimental data for the excitonic mirror case of our device with a 12 nm squaraine layer. (e) Schematic drawing of spectrally selective modification of the cavity round trip phase by an excitonic mirror. (f) Reflection phase from transfer-matrix calculations of a bare DBR (black), DBR with a 1 nm (blue) and 12 nm squaraine layer (red) superimposed with the dielectric function of squaraine taken from Ref. [24].

## CONCLUSION

In our work, we have demonstrated a peculiar regime of strong light-matter coupling of a thin squaraine layer acting as an ultra-thin excitonic mirror embedded in a spectrally tunable optical cavity. For short cavity length, the system features a very large Rabi-splitting, which approaches the transition to the ultrastrong coupling regime. Strikingly, the collapse of the Rabi-splitting features substantial modifications to the 'standard' scenario, and most notably, a merging of resonances of different mode index in the long cavity limit. This effect is assigned to the excitonic mirror behavior, which selectively switches the lower cavity section from dielectric to metallic behavior for a specific range of energies, introducing a 2π phase change across the exciton resonance. Since excitonic mirrors have been shown to be highly tunable [33] and potentially switchable on ultra-fast time scales, our result outlines an exciting perspective towards room-temperature on-chip photonic and polaritonic applications where the explicit mirror-like character of excitons play the role of active photonic elements.

## AUTHOR INFORMATION

### Corresponding Author

*m.esmann@uni-oldenburg.de

### Author Contributions

The manuscript was written through contributions of all authors. All authors have given approval to the final version of the manuscript.

## ACKNOWLEDGMENT

The authors acknowledge support by the German research foundation (DFG) via the project SCHN1376 13.1. Funded in part by the Deutsche Forschungsgemeinschaft (DFG, German Research Foundation) under Germany's Excellence Strategy - EXC 3051/1 „NaviSense" – project number 533653176. Financial support by the Niedersächsisches Ministerium für Wissenschaft and Kultur ("DyNano" and "ElLiKo") is gratefully acknowledged. MFS and AL are obliged to the DFG GRK 2591 "Template-designed Organic Electronics – TIDE" for financial support. MFS thanks the Manchot Foundation for a doctoral scholarship. M.E. acknowledges support by the University of Oldenburg through a Carl von Ossietzky Young Researchers' Fellowship. C.S. acknowledges support by the European Research Council (ERC) within the project 'Dual-Twist' (grant agreement number 101170213).

# Supporting Information

# Strong coupling regimes of an organic exciton mirror in a microcavity

*Christoph Bennenhei[1], Lukas Lackner[1], Moritz Gittinger[1], Falk Eilenberger[2], Marvin F. Schumacher[3], Arne Lützen[3], Ivan Shelykh[4], Christoph Lienau[1], Martin Esmann[1], Christian Schneider[1]*

[1]*Institute of Physics, Fakultät V, Carl von Ossietzky Universität Oldenburg, 26129 Oldenburg, Germany*

[2]*Institute of Applied Physics, Abbe Center of Photonics, Friedrich Schiller University Jena, 07743 Jena, Germany; Fraunhofer-Institute for Applied Optics and Precision Engineering IOF, 07743 Jena, Germany; Max-Planck-School of Photonics, 07743 Jena, Germany*

[3]*Kekulé-Institute for Organic Chemistry and Biochemistry, University of Bonn, 53121 Bonn, Germany*

[4]*University of Iceland, Science Institute, Reykjavik, Iceland*

## S1: Experimental setup

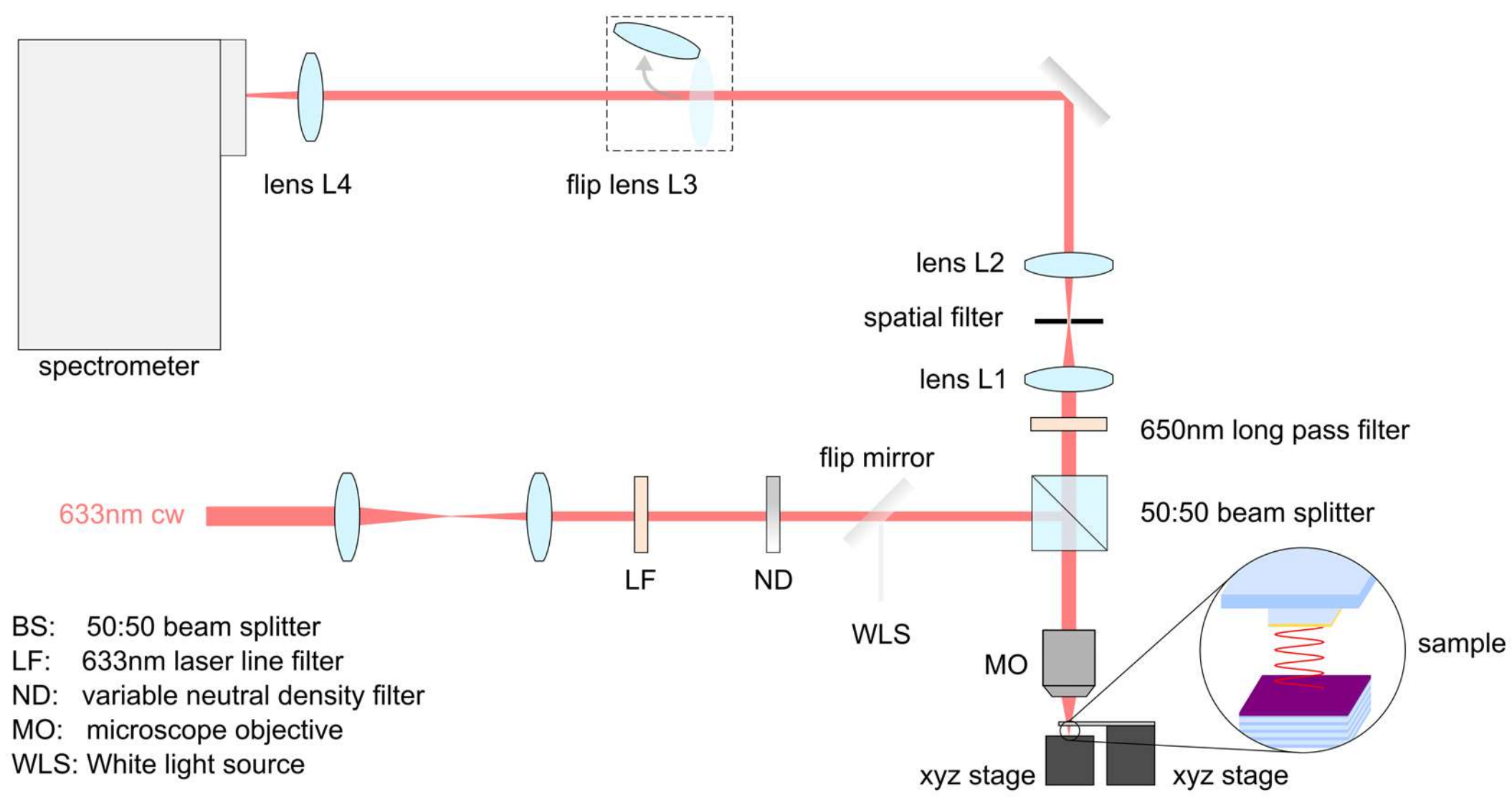

*Figure S1: Experimental setup used for photoluminescence and momentum-resolved white light reflectivity measurements. The open cavity (lower right corner) consists of two independently controlled xyz-piezo stages each supporting one of the two cavity mirrors. Momentum-resolved white light (Thorlabs SLS301) reflectivity spectra are recorded by imaging the back focal plane of the microscope objective (MO, Mitutoyo Plan Apo NIR HR 0.65 NA, 50x) onto the entrance slit of the spectrometer (Andor SR-500i-A-SIL with an Andor iKon-M 934 CCD camera) in a 4f configuration when lens L3 is in place (focal lengths: MO = 4 mm, f1 = 300 mm, f2 = 200 mm, f3 = 300 mm f4 = 400 mm).*

## S2: Extended Reflectivity data and mode profiles

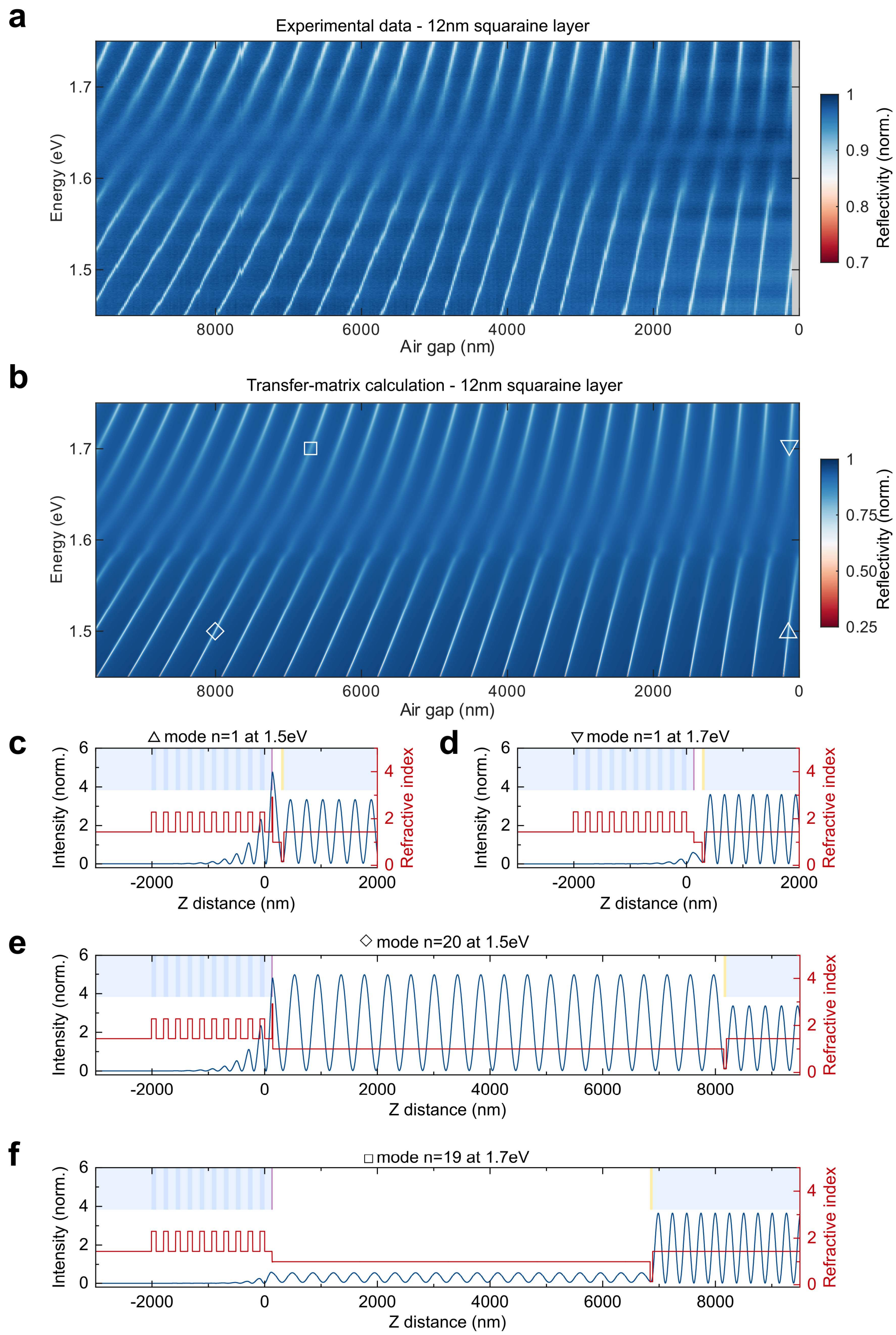

*Figure S2: Strong coupling for a 12 nm thick squaraine film in the open cavity, extended dataset for Fig. 3(a,b) and Fig. 4(d) in the main text. (a) Normal incidence reflectivity spectra ($k_{//}$=0 sections) recorded as a function of the air gap of the cavity. Anti-crossings occur at the exciton energy of 1.59 eV. The gray area marks contact between the mirrors, inhibiting a closer approach. (b) Transfer-matrix calculations corresponding to the experimental data shown in panel (a). Triangles and squares mark resonances for which calculated intensity profiles are plotted in panels (c-f). The intensities are calculated for the fields normalized to the incident field. (c,d) Intensity profile for the polariton mode with index n=1, lower (c) and upper (d) polariton, each with one anti-node in the cavity. The coupling is on the order of the free spectral range leading to characteristic bent modes. (e,f) Intensity profiles for the resonance crossing the exciton as a straight line in panel (b) for an air gap*

*of around 8000 nm, marked with squares. The mode index is n=19 above and n=20 below resonance with intensity profiles differing by one anti-node in the cavity. This is a direct consequence of the spectrally selective ultra-thin excitonic mirror enforcing metallic field boundary conditions on the lower DBR at the exciton resonance.*

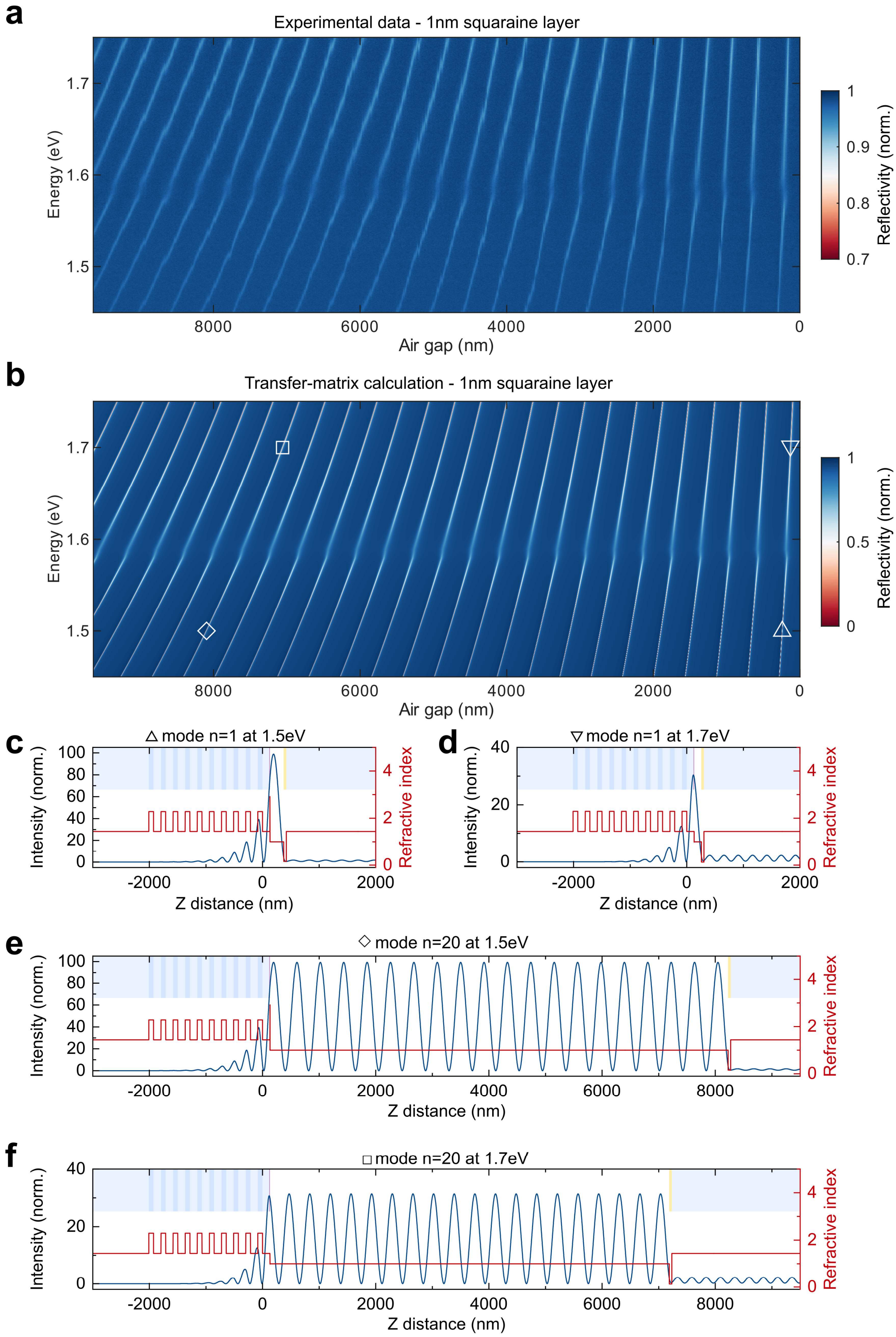

*Figure S3: Strong coupling for a 1 nm thick squaraine film in the open cavity, extended dataset for Fig. 4(b) in the main text. (a) Normal incidence reflectivity spectra ($k_{//}$=0 sections) recorded as a function of the air gap of the cavity. Anti-crossings occur at the exciton energy of 1.59 eV. (b) Transfer-matrix calculations corresponding to the experimental data shown in panel (a). Triangles and squares mark resonances for which calculated intensity profiles are plotted in panels (c-f). The intensities are calculated for the fields normalized to the incident field. (c,d) Intensity profile for the polariton mode with index n=1, lower (c) and upper (d) polariton, each with one anti-node in the cavity. (e,f) Intensity profiles for the resonance crossing the exciton as a straight line in panel (b) for an air gap of around 8000 nm, marked with squares. The mode index is n=20 above and below resonance with both intensity profiles showing 20 anti-nodes in the cavity.*

## S3: Very long cavity limit

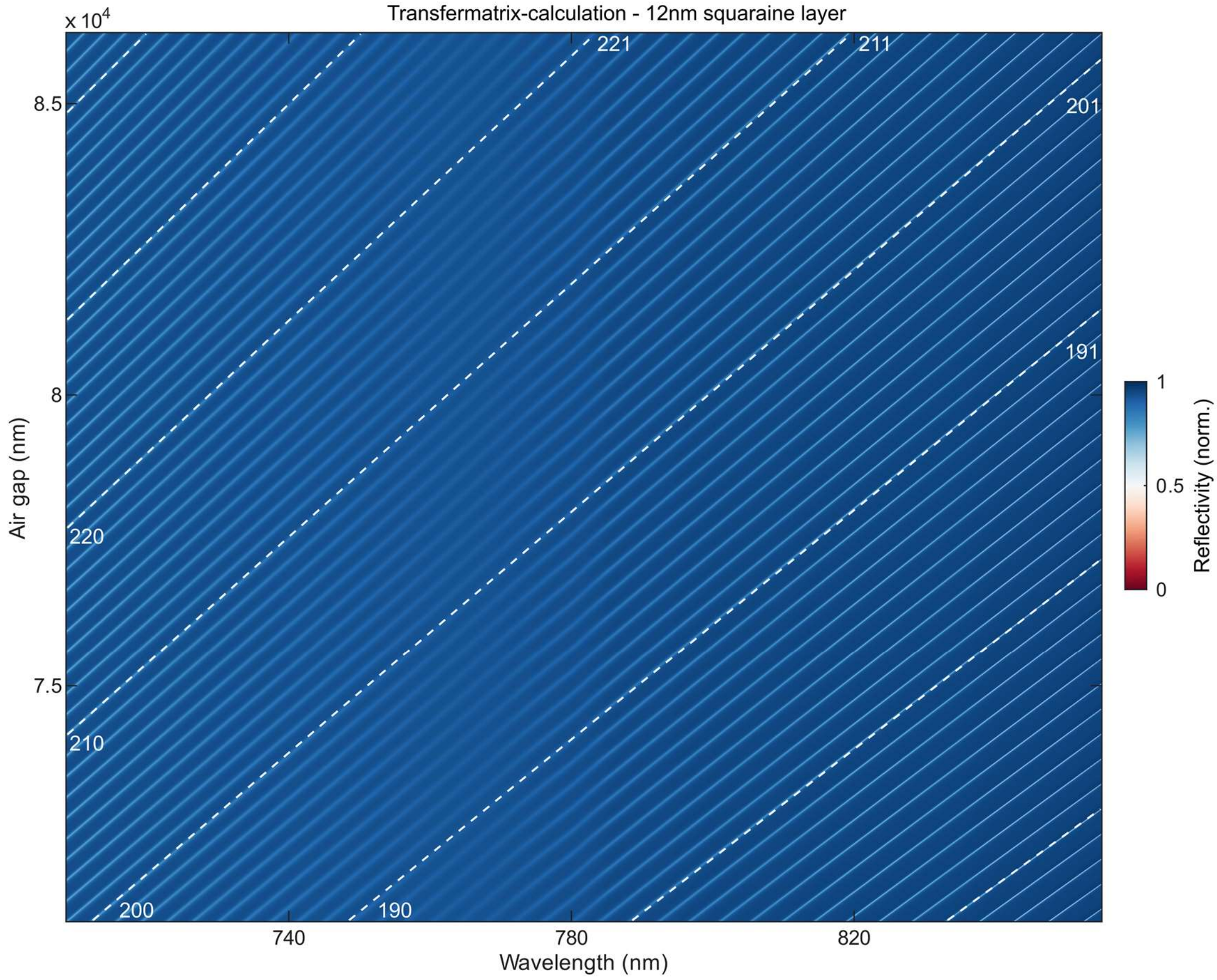

*Figure S4: Transfer-matrix calculations for a 12 nm thick squaraine film in the open cavity, for very large cavity mirror separation (~85 µm). The dashed white lines mark resonances of specific mode orders n and n+1 (white numbers) that smoothly connect across the exciton resonance, due to the excitonic mirror behavior. This finding distinctly differs from the standard weak coupling scenario shown in Figure S5.*

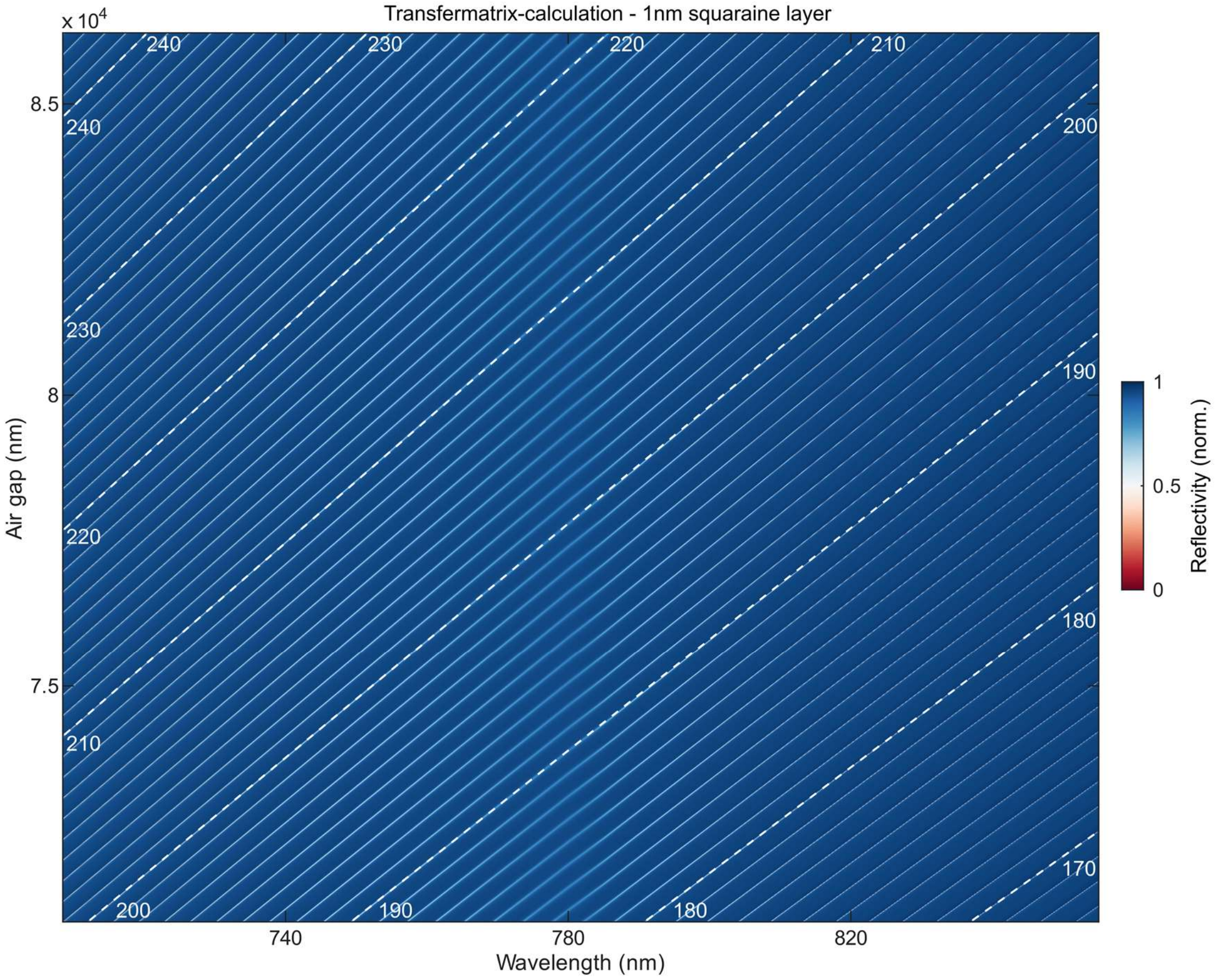

*Figure S5: Transfer-matrix calculations for a 1 nm thick squaraine film in the open cavity, for very large cavity mirror separation (~85 µm). The dashed white lines mark resonances of specific mode orders n (white numbers) that smoothly cross the exciton resonance, as expected in the standard weak coupling scenario.*

## S4: Strong-to-weak coupling transition

The observed linking of modes of different mode order as consequence of the excitonic mirror persists in both the strong and weak coupling limit.

To elaborate this point, we estimate the minimum cavity length for which we can be certain that our system operates in the weak coupling regime. In Figure 3c of the main text, we provide the experimentally measured Rabi-splitting $\Omega$ as a function of cavity length $L_{cav}$ together with a fit following $\Omega = A/\sqrt{L_{cav}}$ for a squaraine film of 12 nm thickness. To get an upper-limit estimate for the cavity length at which the weak to strong coupling transition occurs, we assume lower-limit losses $\gamma_{cav}$ for the empty cavity mode and $\gamma_x$ for the exciton resonance. These HWHM values are obtained from Lorentzian fits to the numerically calculated absorption $A_{Sq} = 1 - T - R$ of the squaraine film (with transmission T and reflectivity R) and to the calculated reflectivity of the empty cavity, respectively. We obtain:

Empty cavity: $\gamma_{cav} = 0.34$ meV

12 nm Squaraine film: $\gamma_x = 50.45$ meV

From these linewidths, we calculate the necessary coupling strength g for strong coupling (with a Rabi-splitting $\Omega = 2\sqrt{g^2 - \frac{1}{4}(\gamma_{cav} - \gamma_x)^2}$ ) [1] using either the visibility criterion

$g_{sc} = \frac{1}{2}\sqrt{(\gamma_{cav} - \gamma_x)^2 + \gamma_{cav} \cdot \gamma_x}$ [2] or the strong to weak coupling transition criterion

$g_{sw} = \left|\frac{\gamma_{cav} - \gamma_x}{2}\right|$ [1]. We obtain nearly identical values for the two criteria:

$$g_{sc,12nm} = 25.14 \text{ meV}$$

$$g_{sw,12nm} = 25.05 \text{ meV}$$

By comparison to the fit shown in Fig. 3c of the main text, these values imply that we enter the weak coupling regime for cavity lengths above ~11 µm in case of a 12 nm squaraine film. Therefore, the calculations performed in Figures S4 and S5 above are both deep in the weak coupling regime. This means that the observed linking of modes of different mode order as a consequence of the excitonic mirror persists in both the strong and weak coupling limit.

## S5: Mode profile videos

To further illustrate the continuity of modes across the exciton we performed simulations of the modal intensity distribution tracing individual modes across the exciton resonance. That is, in addition to showing modes at two selected energies in Figures S2(e,f) and S3(e,f), we have calculated intensity profiles energetically very close to the exciton resonance and plotted them in Fig. S6 below. These profiles highlight the smooth transition in mode number as one cavity mode anti-node continuously disappears at the excitonic mirror surface with increasing energy (panel a). This is contrasted by the standard cavity behavior of a thinner film (panel b).

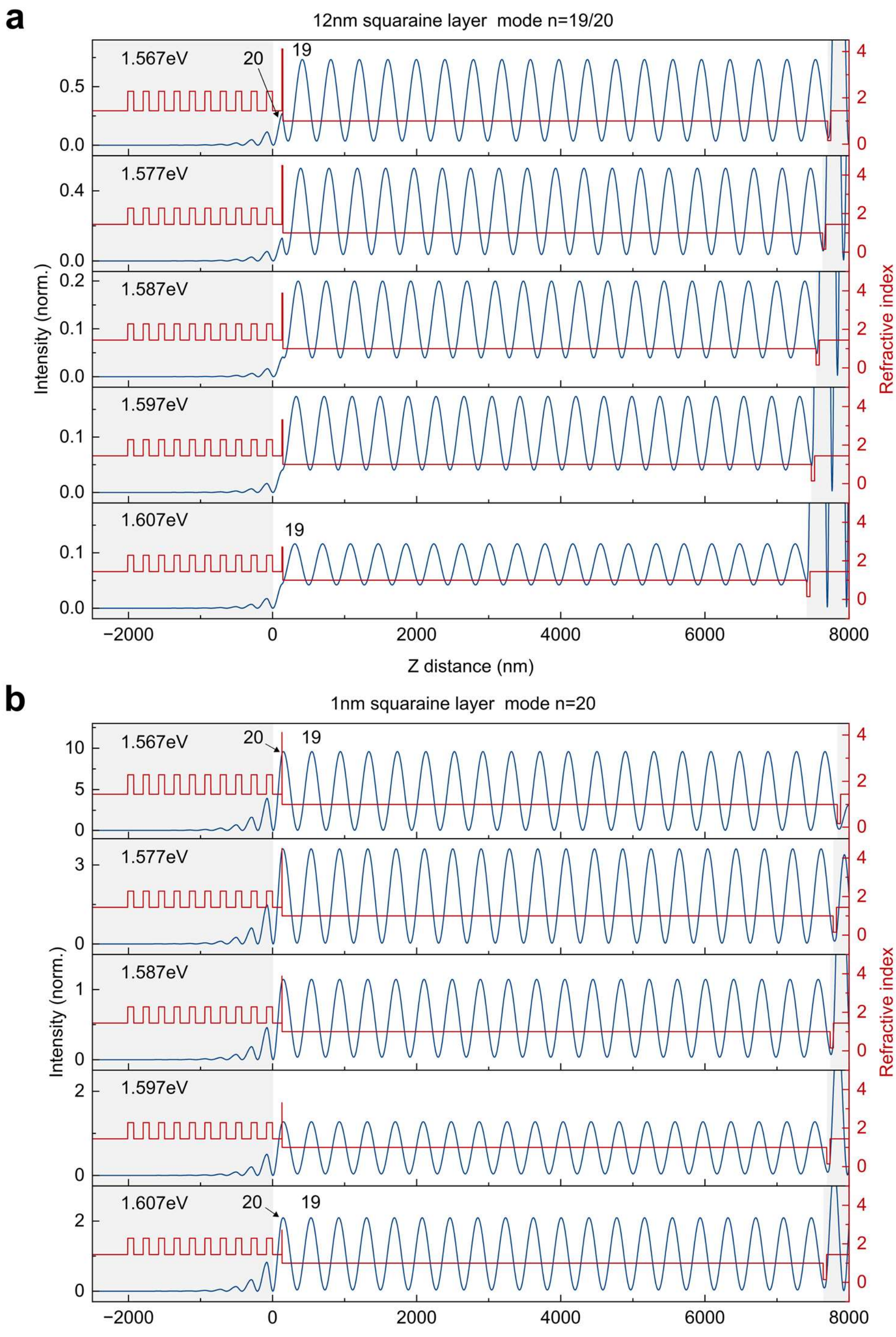

*Figure S6: (a) Intensity profiles in the vicinity of the exciton resonance at 1.59 eV for a 12 nm squaraine film in the cavity calculated from transfer-matrix calculations. For better visibility the areas outside of the cavity volume are highlighted in gray. Note that the last lower refractive index layer of the DBR acts as a spacer for the squaraine layer to be positioned in a field maximum at resonance and thus the cavity volume is made up by the low refractive index spacer, the active material and the air gap. For lower energies, the cavity volume contains 20 anti-nodes of the intensity profile, with the 20th anti-node damped by the excitonic layer. Tuning the cavity to higher energies, the 20th anti-node continuously reduces and merges into the flank of the 19th anti-node when crossing the exciton resonance. This smooth linking of two modes of different mode order is facilitated by the excitonic mirror behavior of the DBR/squaraine system. (b) Calculated intensity profiles for the same energies as in (a), but for a 1 nm squaraine layer. Here, 20 anti-nodes persist in the cavity volume and no linking of modes of different mode order is observed. Y-scales in both panels are optimized for visibility of the field anti-nodes 19 and 20.*

## S6: Excitonic mirror behavior as a function of squaraine film thickness

We explore the excitonic mirror behavior of a squaraine film on a DBR by investigating its spectrally dependent reflection phase as a function of the squaraine layer thickness. That is, we probe the essential property of our excitonic mirror as a function of organic film thickness.

We show calculated reflection phases in Figure S7 below. As described in the main text, adding a squaraine layer to the DBR introduces a change in reflection phase at the exciton resonance. This change increases with rising layer thickness, leading to a π phase jump at a layer thickness of 5.5 nm. This thickness marks the transition from a cavity system with regular strong/weak coupling (like the 1 nm film studied experimentally) to the excitonic mirror regime (like the 12 nm film). Increasing the layer thickness further, the phase change across the exciton resonance persists, but flattens out resulting in an overall change in round-trip phase of $2\pi$.

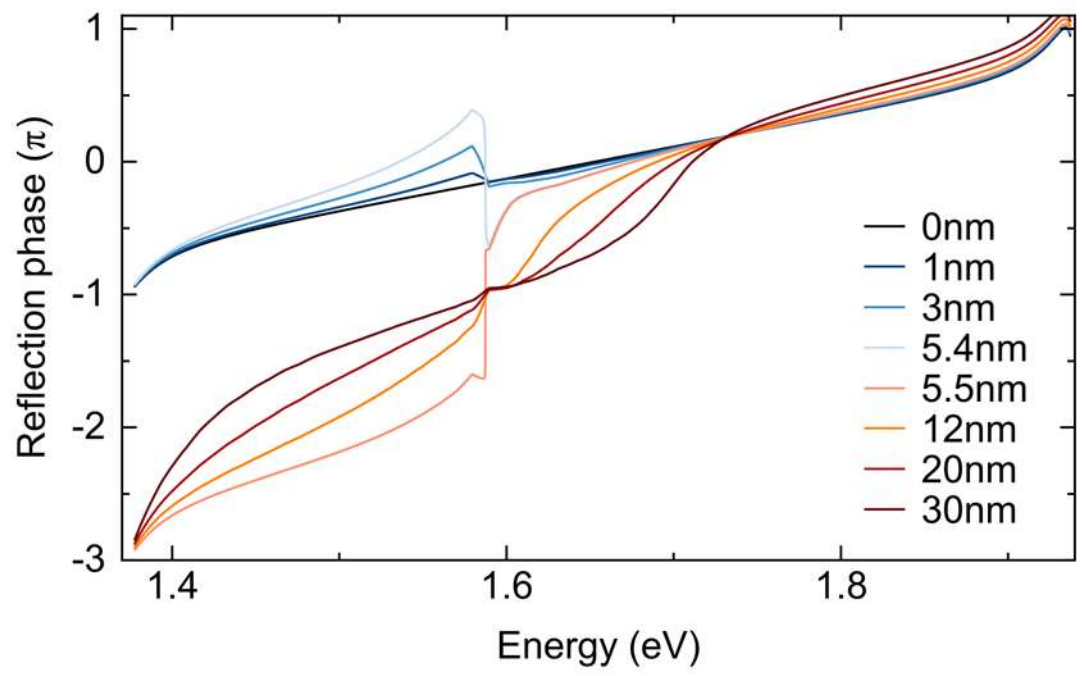

*Figure S7: Spectral reflection phase of squaraine layers with increasing thickness placed on a DBR obtained from transfer matrix calculations. The squaraine layers introduce a change of the reflection phase at the exciton resonance. For layer thicknesses above 5.5 nm a phase occurs that persists for higher layer thicknesses. This thickness thus marks the boundary between a system with regular weak-to-strong coupling transition and the excitonic mirror regime.*

## SI References